\documentclass[runningheads]{svmult}

\usepackage{makeidx}   
\usepackage{graphicx}  
\usepackage{subeqnar}  
\usepackage{multicol}  
\usepackage{physprbb}  
\makeindex             

\begin{document}
\title*{H$_2$ Mid-IR Pure Rotational Emission from Young Stars:
The TEXES/IRTF Survey}

\toctitle{H$_2$ Mid-IR Emission from Young Stars:
The TEXES/IRTF Survey}
\titlerunning{H$_2$ Mid-IR Emission from Young Stars: The TEXES/IRTF Survey}
%
\author{Matthew J. Richter\inst{1,4}
\and John H. Lacy\inst{2,4}
\and Thomas K. Greathouse\inst{2,4}
\and Daniel T. Jaffe\inst{2,4}
\and Geoffrey A. Blake\inst{3}}
\authorrunning{Richter et al.}
%
%
\institute{University of California, Davis CA 95616, USA
\and University of Texas, Austin TX 78703, USA
\and California Institute of Technology, Pasadena CA 91125, USA
\and Visiting Astronomer at the Infrared Telescope Facility, which is 
operated by the University of Hawaii under Cooperative Agreement 
no. NCC 5-538 with the National Aeronautics and Space Administration, 
Office of Space Science, Planetary Astronomy Program.}

\maketitle              

\begin{abstract}
We describe the TEXES survey for mid-IR H$_2$
pure rotational emission from young stars and report early successes.
H$_2$ emission is a potential tracer of warm gas in circumstellar disks.
Three pure rotational lines are available from the ground: the 
J\,=\,3\,$\Rightarrow$\,1,
J\,=\,4\,$\Rightarrow$\,2, and
J\,=\,6\,$\Rightarrow$\,4,
transitions at 17.035\,$\mu$m, 12.279\,$\mu$m, and 
8.025\,$\mu$m, respectively.  
Using TEXES at the NASA IRTF 3m, we are midway through a survey of roughly
30 pre-main-sequence stars.  
To date,
detected lines are all resolved, generally with FWHM$<$10 km/s.  
Preliminary analysis suggests the gas temperatures are between
400 and 800~K.
From the work so far, we conclude that high spectral and
spatial resolution are critical to the investigation of H$_2$ in disks.
\end{abstract}

\section{Introduction}

With the excitement caused by the detection of 
extra-solar planets~\cite{mayor95}, there has been growing emphasis on 
the study of circumstellar disks as the birthplace of diverse
planetary systems.  While great progress has been made in the study of 
circumstellar dust at both near-infrared (e.g.~\cite{lada00}) and 
millimeter wavelengths (e.g.~\cite{looney00}). The observations generally
do not provide information at disk radii comparable to the orbits of planets
in our own solar system.  Furthermore, knowledge of the gas 
is critical to
understanding the formation of Jovian planets.

Circumstellar gas studies are also making tremendous progress.  Most
of the work involves CO rotational lines~\cite{dutrey96} or ro-vibrational
lines~\cite{najita03}, although hot H$_2$O ro-vibrational transitions
are another tool~\cite{carr03}.
While the rotational lines are dominated by gas
at large radii ($\sim$100~AU), the ro-vibrational lines are able to probe 
regions out to $\sim$1~AU.  

An alternative method of looking at circumstellar gas is to study 
H$_2$ directly.  H$_2$ is the most abundant molecule, does not freeze onto dust
grains, and has has electronic transitions in the UV~\cite{herczeg02},
ro-vibrational transitions in the near-infrared~\cite{bary03},  
and rotational transitions in the mid-infrared.
The mid-infrared lines come from
low lying rotational states that
are easily thermalized, are almost
completely unaffected by
extinction, and are optically thin.
With sufficient spectral
resolution, it may be possible to invert the H$_2$ line
profile and derive the gas temperature in the disk as a function of radius.
Unfortunately, the mid-infrared lines are weak, as they arise from 
quadrupole transitions.
Three low energy rotational lines of H$_2$ are available from ground-based 
telescopes:
the J\,=\,3\,$\Rightarrow$\,1 
at 17.035\,$\mu$m (also called the 0-0 S(1) line);
the J\,=\,4\,$\Rightarrow$\,2 at 12.279\,$\mu$m (the 0-0 S(2) line);
and the
J\,=\,6\,$\Rightarrow$\,4 at 8.0251\,$\mu$m (the 0-0 S(4) line).

Thi et al.~\cite{thi01a,thi01b} used the Infrared Space
Observatory (ISO) to survey young stars for circumstellar H$_2$.  Their results
suggested the gas was cold, T$\sim$100-200~K, and that significant reservoirs
of H$_2$ persisted even around main sequence stars with no apparent 
CO.  Richter et al.~\cite{richter02}, 
using ground-based observations of the 
J\,=\,3\,$\Rightarrow$\,1 transition, showed ISO could 
not have been detecting disk gas.  

In order to test the usefulness of H$_2$ as a probe of disk gas,
we began the survey of young stars
described here to examine the frequency of disk H$_2$ emission.
In the following sections, we will describe the survey (Section 2), some
early general results (Section 3), and finish with a short discussion of 
our preliminary results along with guidelines for future 
attempts to utilize the mid-IR H$_2$ emission lines
(Section 4).
More detailed discussions of the survey are 
in preparation for future publication.

\section{Observations and the Survey Strategy}

\begin{figure}[t]
\begin{center}
\includegraphics[width=.85\textwidth]{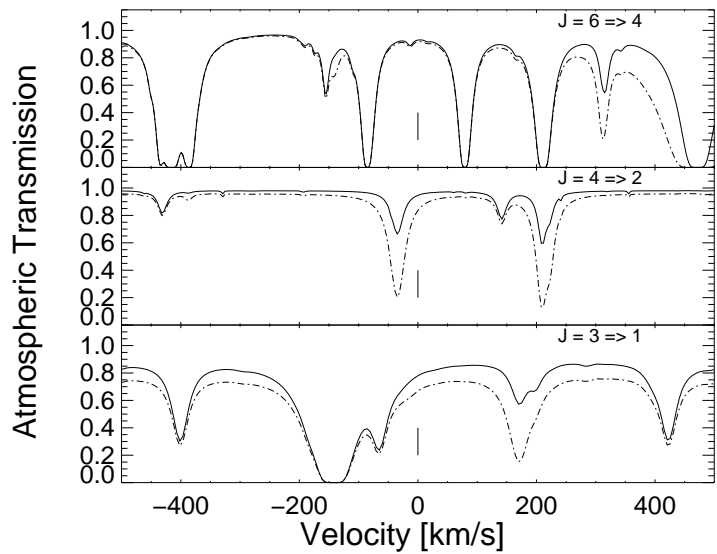}
\end{center}
\caption[]{Model atmospheric transmission for Mauna Kea at the spectral
settings corresponding to the three H$_2$ lines available from the ground.
In each panel, the solid line represents 1.0~mm precipitable water vapor
and the dash-dot line represents 4.0~mm precipitable water vapor.}
\label{fig1}
\end{figure}

Our H$_2$ survey is made possible by the high spectral
and spatial resolution provided by TEXES, the
Texas Echelon-cross-Echelle Spectrograph~\cite{lacy02}.
TEXES is a mid-infrared, cross-dispersed spectrograph with 
resolving power $\sim$80,000
at 12.3\,$\mu$m.
On a 3~meter telescope such as NASA's Infrared Telescope Facility 
(IRTF), TEXES provides the sensitivity to match or surpass ISO for 
observations of unresolved sources with narrow lines.  
The high spectral resolution not
only provides optimal sensitivity to narrow lines, but also helps to separate
atmospheric features from those of astronomical interest.  This is
particularly important for observing H$_2$ lines (Fig~\ref{fig1}).
In addition, the spatial resolution with TEXES approaches the 
diffraction-limit (0.85$^{\prime\prime}$ at 
12.3\,$\mu$m on a 3~meter telescope) giving us access to spatial
clues for understanding the emission.

\begin{figure}[t]
\begin{center}
\includegraphics[width=.85\textwidth]{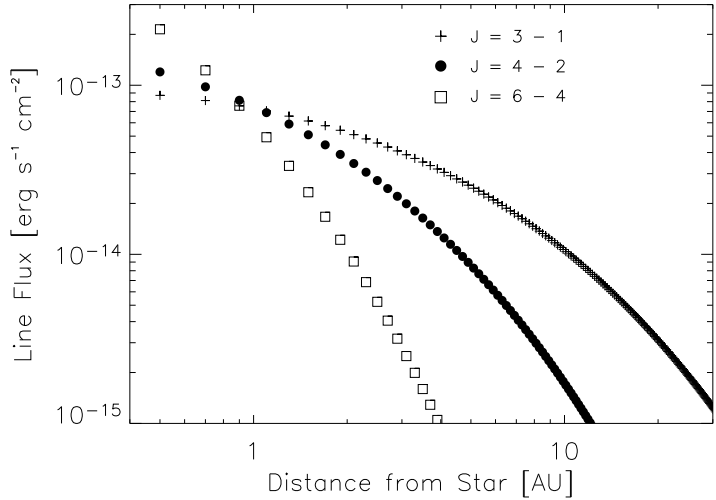}
\end{center}
\caption[]{An estimate of the line flux from the three mid-infrared 
rotational lines
of H$_2$ available from the ground assuming the gas is the same temperature
as a hot disk surface
layer composed of small grains~\cite{chiang97}.  
At each radial step, we calculate the line flux
from a gas mass of 10~M$_\oplus$ at a distance of 140 pc.  In one hour,
TEXES observations on the IRTF can reach 3$\sigma$ detection limits 
$<10^{-14}$\,erg/s/cm$^2$.}
\label{fig2}
\end{figure}

Based on our previous work~\cite{richter02}, we 
expected 
to reach line flux limits comparable to or better than those obtained
with ISO in roughly one hour.  
This realization, along with the non-detection of H$_2$ in 
ISO sources, prompted 
our survey of roughly 30 young stars.
Since no one knows 
the characteristic temperature of circumstellar
H$_2$ emission,
we concentrate on 
J\,=\,3\,$\Rightarrow$\,1 and J\,=\,4\,$\Rightarrow$\,2.  
The ratio of these two lines is sensitive to
gas temperatures in the range 200-600~K.
Figure~\ref{fig2} shows that 10~M$_\oplus$ of H$_2$ gas can readily
be detected at a distance of 140~pc for the case of gas in equilibrium with
a radiatively heated dust surface layer composed of small 
grains~\cite{chiang97}.
For all sources where we detect line emission,
we will also observe J\,=\,6\,$\Rightarrow$\,4.  The third
line provides additional leverage on the gas temperature, up to $\sim$1000~K,
as well as a
glimpse at the ortho:para ratio for H$_2$.  Finally, while the 
observations of the two
primary lines are sensitive to the amount of water 
in our own atmosphere,
the J\,=\,6\,$\Rightarrow$\,4 line observations are relatively unaffected
by terrestrial water vapor (Fig~\ref{fig1}).

We currently concentrate on sources with relatively
strong mid-infrared continuum fluxes to allow guiding through
TEXES.  We use the IRTF visible camera for source acquisition and
offset guiding whenever possible, but
long integrations on weak mid-infrared sources
raise concerns about the pointing and tracking efficiency.
Unfortunately, requiring a strong mid-infrared
continuum biases the survey toward the brightest sources and
diminishes its general applicability.  
Consequently, our survey targets include more higher mass
Herbig Ae stars and fewer weak-lined T~Tauri stars and debris disks
than desired.  A larger telescope or an internal
guider would improve our survey.

In making our observations, we concentrate on detecting material within
$\sim$200~AU of  
the central source.  Typical instrumental setups are given in 
Table~\ref{tab1}.
To remove background emission, we nod the telescope
so the object is always on the slit.  This renders us insensitive
to emission on scales larger than $\sim$5$^{\prime\prime}$.  We oriented
the slit north-south for all observations.

\begin{table}
\caption{Typical TEXES Observing Parameters}
\begin{center}
\renewcommand{\arraystretch}{1.4}
\setlength\tabcolsep{5pt}
\begin{tabular}{cccc}
\hline\noalign{\smallskip}
Setting & $\mathrm{R}\equiv\lambda / \Delta \lambda$ & Slit Width & 
  Slit Length \\
\noalign{\smallskip}
\hline
\noalign{\smallskip}
 J\,=\,3\,$\Rightarrow$\,1 & 60,000 & 2$^{\prime\prime}$ & 
10.5$^{\prime\prime}$ \\
 J\,=\,4\,$\Rightarrow$\,2 & 90,000 & 1.4$^{\prime\prime}$ & 
7.5$^{\prime\prime}$ \\
 J\,=\,6\,$\Rightarrow$\,4 & 90,000 & 1.4$^{\prime\prime}$ & 
7.5$^{\prime\prime}$ \\
\hline
\end{tabular}
\end{center}
\label{tab1}
\end{table}

%

\section{Results}

\begin{figure}[t]
\begin{center}
\includegraphics[width=.85\textwidth]{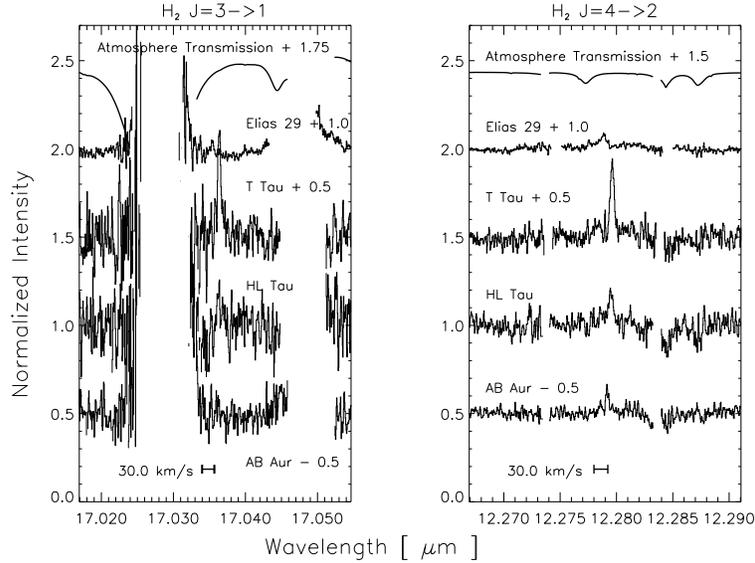}
\end{center}
\caption[]{Portions of the spectra 
recorded at 17 and 12~$\mu$m from sources with 
clear J~=~4~$\Rightarrow$~2 detections.  We show normalized
spectra offset for clarity with wavelength scale as observed.
Both panels include a telluric transmission spectrum, although
the offset is different because of the low transmission at 17~$\mu$m.
Gaps in the spectral coverage  result from the spectral order
being larger than the detector.
}
\label{fig3}
\end{figure}

Roughly a third of the 21 stars observed so far show 
evidence of H$_2$ emission at one or more settings.  No clear, 
symmetric, double-peaked line profiles indicative of disk emission
have been seen.  While almost all the detected lines are clearly 
resolved, most lines are relatively
narrow with a typical FWHM $\approx$10~km/s and a single central peak.  
With a high resolution 
instrument such as TEXES, we are naturally biased toward narrow lines
that concentrate the flux in a few resolution elements, but we do not feel
this bias is the reason we see only narrow lines.
Fluxes are on the order of
a few\,$\times$\,10$^{-14}$
ergs/s/cm$^2$ with equivalent widths of 0.5 to 5 km/s.  
Figure~\ref{fig3} give examples 
of a few of the clear detections.
The T~Tau system has the highest line
flux and line-to-continuum ratio.
In general, the 
J~=~4~$\Rightarrow$~2 lines are clearer than the 
J~=~3~$\Rightarrow$~1 lines.  While the instrument sensitivity
and telluric transmission are better at 12~$\mu$m than at 17~$\mu$m, 
the population of J~=~4 requires relatively hotter gas than we had
naively anticipated.  

In almost every case, the line emission is coincident with the mid-IR 
continuum.  However, in the case of T~Tau, a fairly complicated multiple
system, the line emission comes from roughly 0.5$^{\prime\prime}$ south of
the mid-infrared continuum source.  

\section{Discussion}

By taking the measured line fluxes, we can use the line ratios to determine
an H$_2$ temperature and mass.  For the case of T~Tau, as will be described
in a future paper, we determine a gas temperature of $\sim$400-800~K and
a mass of $\sim$5-10~M$_\oplus$.  Errors in the flux 
determinations at the three wavelength settings will have a large effect
on the final temperature and mass determinations.  We have tried to 
account for these effects in the ranges given.

Given the narrow line widths we see and the centrally-peaked nature of
the line profiles, it is unlikely our first detections come from disk 
material close to the central star or 
gas shock-excited and associated with 
a bipolar outflow.  The material close to
the star in a Keplerian disk provides the greatest velocity extent.
If the inner regions produce no emission, the
intrinsic line shape,
when convolved with the instrumental profile, would more closely match a 
single centrally-peaked profile.  
Line-of-sight effects -- a disk inclined to the line of sight or a shock
traveling perpendicular to the line of sight -- could also result in narrow 
line widths.  
Disks with their rotation axes within roughly 20 degrees of the line of
sight will show a single centrally-peaked profile.  
Naturally, higher signal-to-noise observations would help in the study of the
line profiles.

As mentioned above, 
flux determinations through our spectrograph
slit are a concern since calibration errors result 
in large allowed temperature and mass ranges.
Ideally, the line strength would be fixed to the mid-IR continuum which can
be accurately determined from imaging.  Unfortunately, even this 
method can introduce errors if the source varies in the mid-infrared, such as 
the T~Tau system~\cite{ghez91}.  Based on our experiences, future 
efforts for studying H$_2$ disk emission should be very careful about
establishing good fluxes.

Finally, we note that the TEXES spectra show the importance of both high
spatial and spectral resolution.  The spatial offset between continuum and
line emission, such as seen in the T~Tau system, 
requires spatial resolution $\sim$1$^{\prime\prime}$.  
Since the equivalent widths of the emission detected so far is less than
$<$5 km/s, instruments with R$<$10,000 will require very high signal-to-noise
ratios in order to detect the lines and will be have a limited ability  
to examine the velocity structure in what appears to be the
small fraction of sources with broad lines. 

Support for this work comes from a Texas Advanced Research Program grant 
TARP 00365-0473-1999.  MJR acknowledges grant USRA 8500-98-008 through 
the SOFIA program and NSF grant AST-0307497. This research has made use of 
NASA's Astrophysics Data System and the SIMBAD database operated at the 
Centre de Données astronomiques de Strasbourg, France.

\end{document}